\documentclass[prl,showpacs,twocolumn]{revtex4}%
\input{psfig.sty}
\usepackage{amsmath}
\usepackage{graphicx}
\usepackage{amsfonts}
\usepackage{amssymb}

\begin{document}
\title{Deterministic single photons via conditional quantum evolution}
\date{\today }
\author{D. N. Matsukevich, T. Chaneli\`{e}re, S. D. Jenkins, S.-Y. Lan, T. A. B. Kennedy, and A. Kuzmich}
\affiliation{School of Physics, Georgia Institute of Technology, Atlanta, Georgia 30332-0430}
\begin{abstract}
A source of deterministic single photons is proposed and
demonstrated by the application of a measurement-based feedback
protocol to a heralded single photon source consisting of an
ensemble of cold rubidium atoms. Our source is stationary and
produces a photoelectric detection record with sub-Poissonian
statistics.
\end{abstract}
\maketitle

Quantum state transfer between photonic- and matter-based quantum
systems is a key element of quantum information science,
particularly of quantum communication networks. Its importance is
rooted in the ability of atomic systems to provide excellent
long-term quantum information storage, whereas the long-distance
transmission of quantum information is nowadays accomplished using
light. Inspired by the  work of Duan {\it et al.} \cite{duan},
emission of non-classical radiation has been observed in
first-generation atomic ensemble experiments \cite{kuzmich}.

In 2004 the first realization of coherent quantum state transfer
from a matter qubit onto a photonic qubit was achieved
\cite{matsukevich}. This breakthrough laid the groundwork for
several further advances towards the realization of a long-distance,
distributed network of atomic qubits, linear optical elements and
single-photon detectors
\cite{matsukevich1,chaneliere,revival,matsukevich2,cascade}. A
seminal proposal for universal quantum computation with a similar
set of physical resources has also been made \cite{knill}.

An important additional tool for quantum information science is a
deterministic source of single photons. Previous implementations of
such a source used {\it single emitters}, such as quantum dots
\cite{michler,santori}, color centers \cite{brouri,kurtsiefer},
neutral atoms \cite{kuhn,grangier3}, ions \cite{lange}, and
molecules \cite{moerner}. The measured efficiency $\eta_D$ to detect
a single photon per trial with these sources is typically less than
1\%, with the highest reported measured value of about 2.4\%
\cite{kuhn}, to our knowledge.

\begin{figure}[htp]
\begin{center}
\vspace{-0.36cm} \leavevmode \psfig{file=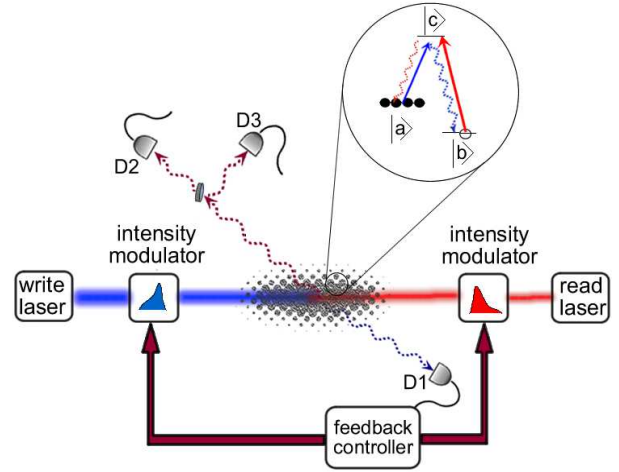,height=2.5in}
\end{center}
\vspace{-0.65cm} \caption{ Schematic of experimental setup, with the
inset showing the atomic level scheme (see text).}\label{TQ}
\end{figure}

We propose a deterministic single photon source based on an {\it
ensemble of atomic emitters}, measurement, and conditional quantum
evolution. We report the implementation of this scheme using a cold
rubidium vapor, with a measured efficiency $\eta _D$ $\approx
1-2\%$. In common with the cavity QED system, our source is suitable
for reversible quantum state transfer between atoms and light, a
prerequisite for a quantum network. However, unlike cavity QED
implementations \cite{kuhn}, it is unaffected by intrinsically
probabilistic single atom loading. Therefore, it is stationary and
produces a photoelectric detection record with truly sub-Poissonian
statistics.

The key idea of our protocol is that a single photon can be
generated at a predetermined time if we know that the medium
contains an atomic excitation. The presence of the latter is
heralded by the measurement of a scattered photon in a {\it write}
process. Since this is intrinsically probabilistic, it is necessary
to perform independent, sequential {\it write} trials before the
excitation is heralded. After this point one simply waits and reads
out the excitation at the predetermined time. The performance of
repeated trials and heralding measurements represents a conditional
feedback process and the duration of the protocol is limited by the
coherence time of the atomic excitation. Our system has therefore
two crucial elements: (a) a high-quality probabilistic source of
heralded photons, and (b) long atomic coherence times. We note that
related schemes using parametric down-conversion have been discussed
\cite{kwiat-franson}.

Heralded single photon sources are characterized by mean photon
number $\langle \hat n \rangle \ll 1 $, as the unconditioned state
consists mostly of vacuum \cite{hong,grangier}. More importantly, in
the absence of the heralding information the reduced density
operator of the atomic excitation is thermal \cite{walls}. In
contrast, its evolution conditioned on the recorded measurement
history of the signal field in our protocol, ideally results in a
single atomic excitation. However, without exception all prior
experiments with atomic ensembles did not have sufficiently long
coherence times to implement such a feedback protocol
\cite{matsukevich,matsukevich1,matsukevich2,chaneliere,revival,kuzmich,eisaman,balic,black}.
In earlier work quantum feedback protocols have demonstrated control
of non-classical states of light \cite{smith} and motion of a single
atom \cite{fischer} in cavity QED.

We first outline the procedure for heralded single photon
generation. A schematic of our experiment is shown in Fig.~1. An
atomic cloud of optical thickness $\approx 7$ is provided by a
magneto-optical trap (MOT) of $^{85}$Rb. The ground levels
$\{|a\rangle;|b\rangle \}$ correspond to the
$5S_{1/2},F_{a,b}=\{3,2\}$ hyperfine levels, while the excited level
$|c\rangle$ represents the $\{5P_{1/2},F_c=3\}$ level of the $D_1$
line at 795 nm. The experimental sequence starts with all of the
atoms prepared in level $|a\rangle $. An amplitude modulator
generates a linearly polarized 70 ns long {\it write} pulse tuned to
the ${|a\rangle \rightarrow |c\rangle }$ transition, and focused
into the MOT with a Gaussian waist of about $430$ $\mu$m. We
describe the {\it write} process using a simple model based on
nondegenerate parametric amplification. The light induces
spontaneous Raman scattering via the ${|c\rangle \rightarrow
|b\rangle }$ transition. The annihilation of a {\it write} photon
creates a pair of excitations: namely a signal photon and a
quasi-bosonic collective atomic excitation \cite{duan}. The
scattered light with polarization orthogonal to the {\it write}
pulse is collected by a single mode fiber and directed onto a single
photon detector D1, with overall propagation and detection
efficiency $\eta_s$. Starting with the correlated state of signal
field and atomic excitation, we project out the vacuum from the
state produced by the write pulse using the projection operator
:$\hat{1} - e^{-\hat{d}^{\dag}\hat{d}}$:, where
$\hat{d}=\sqrt{\eta_s}\hat{a}_s + \sqrt{1-\eta_s}\hat{\xi}_s$,
$\hat{a}_s$ is the detected signal mode, and $\hat{\xi}_s$ is a
bosonic operator accounting for degrees of freedom other than those
detected. Tracing over the signal and all other undetected modes, we
find that the density matrix for the atomic excitation $A$
conditioned on having at least one photoelectric detection event is
given by \cite{prob}
\begin{equation}
\rho_{A|1}=\frac{1}{p_1}\sum_{n=1}^{\infty}\frac{\tanh ^{2n}\chi
}{\cosh^{2}\chi}\left( 1-\left(  1-\eta_{s}\right)  ^{n}\right)
|n\rangle \langle n|,
\end{equation}
where $p_{1} \ll 1 $ is the probability of a signal photoelectric
detection event per {\it write} pulse, and the interaction parameter
$\chi$ is given in terms of $p_{1}$ and $\eta_{s}$ by
\begin{equation}
\sinh^{2}\chi =p_{1}/[\eta_{s}\left(  1-p_{1}\right)  ],
\end{equation}
where $|n\rangle \equiv \hat{A}^{\dagger n}|0\rangle /\sqrt{n!} $,
and $|0\rangle$ is the atomic vacuum. We note that in Eq. (1) there
is zero probability to find $|0\rangle$.

After a storage time $\tau$, a {\it read} pulse of length 80 ns
containing around $3\cdot 10^7$ photons, and with polarization
orthogonal to that of the {\it write} pulse, tuned to the
${|b\rangle \rightarrow |c\rangle }$ transition, illuminates the
atomic ensemble (Fig.~1). Ideally, the {\it read} pulse converts
atomic spin excitations into the idler field emitted on the
${|c\rangle \rightarrow |a\rangle }$ transition. The elastically
scattered light from the write beam is filtered out, while the idler
field  polarization orthogonal to that of the {\it read} beam is
directed into a 50:50 single-mode fiber beamsplitter. Both {\it
write/read} and signal/idler pairs of fields are counter-propagating
\cite{balic}. The waist of the signal-idler mode in the MOT is about
180 $\mu$m.  The two outputs of the fiber beamsplitter are connected
to detectors D2 and D3. Electronic pulses from the detectors are
gated with 120 ns (D1) and 100 ns (D2 and D3) windows centered on
times determined by the {\it write} and {\it read} light pulses,
respectively. Subsequently, the electronic pulses from D1, D2, and
D3 are fed into a time-interval analyzer which records photoelectric
detection events with a 2 ns time resolution.

The transfer of atomic excitation to the detected idler field at
either $Dk$ (k=2,3) is given by a linear optics relation
$\hat{a}_{k} =\sqrt{\eta_{i}\left( \tau \right) /2}\hat
{A}+\sqrt{1-\eta_{i}\left( \tau \right) /2}\hat{\xi}_k \left( \tau
\right)$, where $\hat{a}_{k}$ depends parametrically on $\tau $ and
corresponds to a mode with an associated temporal envelope $\phi
(t)$, normalized so that $\int _{0}^{\infty } dt |\phi (t)|^2=1$,
and $\hat{\xi}_k\left( \tau \right) $ is a bosonic operator which
accounts for coupling to degrees of freedom other than those
detected. The efficiency $\eta_{i}\left( \tau\right)/2  $ is the
probability that a single atomic excitation stored for $\tau$
results in a photoelectric event at $Dk$, and includes the effects
of idler retrieval and propagation losses, symmetric beamsplitter
(factor of 1/2) and non-unit detector efficiency. We start from the
elementary probability density $Q_{k|1}(t_c)$ for a count at time
$t_c$ and no other counts in the interval $[0,t_c)$, $Q_{k|1}(t_c) =
|\phi (t_c)|^2 \langle :\hat a_k^{\dagger } \hat a_k \exp (-\int
_{0}^{t_c} dt |\phi (t)|^2 \hat a_k^{\dagger } \hat a_k):\rangle $
\cite{gardiner}. Using Eq.(1), we then calculate probability
$p_{k|1} \equiv \int _{0}^{\infty } dt Q_{k|1}(t)$ that detector
$Dk$ registers at least one photoelectric detection event. We
similarly calculate the probability $p_{23|1}$ of at least one
photoelectric event occurring at both detectors. These probabilities
are given by
\begin{align}
p_{2|1}\left(  \tau \right)  = p_{3|1}\left(  \tau \right)
=\Pi\left( \eta_{i}\left(
\tau \right)/2; p_{1},\eta_{s}\right),  \\
p_{23|1} \left(  \tau \right)  =p_{2|1}\left(  \tau
\right)+p_{3|1}\left(  \tau \right)-\Pi\left( \eta_{i}\left( \tau
\right); p_{1},\eta_{s}\right),
\end{align}
where we show the explicit dependence on $\tau$. Here $1-\Pi\left(
\eta;p_{1},\eta_{s}\right)$ is given by
$$
\frac{1}{p_{1}}\left(
\frac{1}{1+\eta\sinh^{2}\chi}-\frac{1}{1+\left(  \eta_{s}
+\eta\left(  1-\eta_{s}\right)  \right) \sinh^{2}\chi }\right).
$$

Our conditional quantum evolution protocol transforms a heralded
single photon source into a deterministic one. The critical
requirements for this transformation are higher efficiency and
longer memory time of the heralded source than those previously
reported \cite{matsukevich1,chaneliere}. In Fig.~2 we show the
results of our characterization of an improved source of heralded
single photons. Panel (a) of Fig.~2 shows the measured intensity
cross-correlation function $g_{si} \equiv
[p_{2|1}+p_{3|1}]/[p_2+p_3]$ as a function of $p_1$. Large values of
$g_{si}$ under conditions of weak excitation - i.e., small $p_1$ -
indicate strong pairwise correlations between signal and idler
photons. The efficiency of the signal photon generation and
detection is given by $\eta _s \rightarrow g_{si}p_1$, in the limit
$\sinh^{2}\chi \ll 1$. We have measured $\eta _s \approx 0.08$,
which includes the effects of passive propagation and detection
losses $\epsilon _s$. It is important to distinguish the {\it
measured} efficiency from the {\it intrinsic} efficiency which is
sometimes employed. The intrinsic efficiency of having a signal
photon in a single spatial mode at the input of the single-mode
optical fiber $\eta _s^0 \equiv (\eta _s/\epsilon _s) \approx 0.24$.
We measure $\epsilon _s \equiv \epsilon _s^f \epsilon _s^t \epsilon
_s^d \approx 0.3$ independently using coherent laser light, where
the fiber coupling efficiency $\epsilon _s^f \approx 0.7$, optical
elements transmission $\epsilon _s^t \approx 0.85$, and the
detection efficiency $ \epsilon _s^d \approx 0.55$. The measured
efficiency of the idler photon detection is $\eta _i \rightarrow
g_{si}(p_{2}+p_{3}) \approx 0.075$. Here $p_2$ and $p_3$ are defined
by expressions analogous to Eq. (2). Similarly, the intrinsic
efficiency for the idler field $\eta _i^0 \equiv (\eta _i/\epsilon
_i)\approx 0.34$, where we measure $\epsilon _i \equiv \epsilon _i^f
\epsilon _i^t \epsilon _i^d \approx 0.22$, with $\epsilon _i^f
\approx 0.75$, $\epsilon _i^t\approx 0.59$, and $\epsilon
_i^d\approx 0.55$. The reported values of $\eta _s \approx 0.08$ and
$\eta _i \approx 0.075$ represent slight improvements on the
previous highest measured efficiencies in atomic ensemble
experiments of $0.04 -0.07$ \cite{chaneliere,matsukevich2}.

The quality of the heralded single photons produced by our source is
assessed using the procedure of Grangier {\it et al.}, which
involves a beamsplitter followed by two single photon counters, as
shown in Fig.~1 \cite{grangier}. An ideal single-photon input to the
beamsplitter results in photoelectric detection at either D2 or D3,
but not both. An imperfect single photon input will result in strong
anticorrelation of the coincidence counts. Quantitatively, this is
determined by the anticorrelation parameter $\alpha$ given by the
ratio of various photoelectric detection probabilities measured by
the set of detectors D1,D2 and D3: $ \alpha
=p_{23|1}/(p_{2|1}p_{3|1}). $ Classical fields must satisfy a
criterion $\alpha \geq 1$ based on the Cauchy-Schwarz inequality
\cite{grangier}. For an ideally prepared single photon state $\alpha
\rightarrow 0$. Panel (b) shows the measured values of $\alpha $ as
a function of $p_1$, with $\mbox{min} \{\alpha \} = 0.012 \pm 0.007$
representing a ten-fold improvement on the lowest previously
reported value in atomic ensembles \cite{chaneliere}.

\begin{figure}[btp]
\begin{center}
\vspace{-0.0cm}\leavevmode
\psfig{file=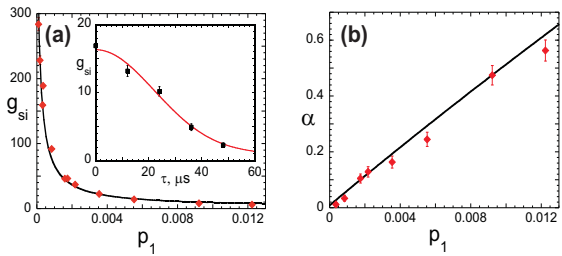,height=1.3in,width=2.9in}
\end{center}
\vspace{-0.5cm} \caption{ Correlation functions $g_{si}$ (panel (a))
and $\alpha $ (panel (b)) as a function of $p_1$, taken at $\tau =
80$ ns. The solid lines are based on Eqs.(3,4), with addition of a
nearly-negligible background contribution, as in Ref.
\cite{chaneliere}. The inset shows normalized signal-idler intensity
correlation function $g_{si}$ as a function of the storage time
$\tau $. The full curve is a fit of the form $1+ B\exp (-\tau
^2/\tau _c ^2)$ with $B= 16$ and the collapse time $\tau _c = 31.5$
$\mu$s as adjustable parameters.} \label{TQ}
\end{figure}

In order to evaluate the atomic memory coherence time $\tau _c$, we
measure $g_{si}$ as a function of the storage time $\tau $, inset of
Fig.~2(a). To maximize $\tau _c$, the quadrupole coils of the MOT
are switched off, with the ambient magnetic field compensated by
three pairs of Helmholtz coils \cite{matsukevich1}. The measured
value of $\tau _c \approx 31.5$ $\mu$s, a three-fold improvement
over the previously reported value, is limited by dephasing of
different Zeeman components in the residual magnetic field
\cite{chaneliere,revival}.

The long coherence time enables us to implement a conditional
quantum evolution protocol. In order to generate a single photon at
a predetermined time $t_p$, we initiate the first of a series of
trials at a time $t_p-\Delta t$, where $\Delta t$ is on the order of
the atomic coherence time $\tau_c$. Each trial begins with a {\it
write} pulse. If D1 registers a signal photoelectric event, the
protocol is halted. The atomic memory is now armed with an
excitation and is left undisturbed until the time $t_p$ when a {\it
read} pulse converts it into the idler field. If D1 does not
register an event, the atomic memory is reset to its initial state
with a cleaning pulse, and the trial is repeated. The duration of a
single trial $t_0 = 300$ ns.  If D1 does not register a heralding
photoelectric event after $N$ trials,  the protocol is halted 1.5
$\mu$s prior to $t_p$, and any background counts in the idler
channel are detected and included in the measurement record.

Armed with Eqs. (3) and (4), we can calculate the unconditioned
detection and coincidence probabilities for the complete protocol.
The probability that the atomic excitation is produced on the
$j^{\text{th}}$ trial is $p_{1}\left( 1-p_{1}\right) ^{j-1}$. This
excitation is stored for a time $(N-j)t_0$ before it is retrieved
and detected, $N=\Delta t/t_{0}$ is the maximum number of trials
that can be performed in the protocol (we ignore the 1.5 $\mu$s
halting period before the read-out).

One can express the probability of a photoelectric event at $Dk$
$\left( k=2,3\right)$, $P_{k}$, and the coincidence probabilities
$P_{23}$ in terms of the conditional probabilities of Eqs. (3) and
(4),
\begin{equation}
P_{\mu }  =p_{1}\sum_{j=1}^{N}\left(  1-p_{1}\right)
^{j-1}p_{\mu|1}\left( \Delta t -jt_{0}\right),
\end{equation}
$\mu =2, 3, 23$. In the limit of infinite atomic coherence time and
$N \rightarrow \infty $,  $P_{\mu } \rightarrow p_{\mu|1}$. Hence,
if the memory time is sufficiently long for an adequate number of
trials, the protocol ideally results in deterministic preparation of
a single atomic excitation, which can be converted into a single
photon at a desired time. Consistent with Fig.~2(a) inset, we assume
a combined retrieval-detection efficiency that decays as a Gaussian
function of storage time, $ \eta_{i}\left( \tau \right)
=\eta_{i}(0)e^{ -( \tau /\tau _c ) ^{2}} $, where $\tau _c$ is the
atomic spin-wave coherence time.
\begin{figure}[btp]
\begin{center}
\vspace{-0.0cm} \leavevmode
\psfig{file=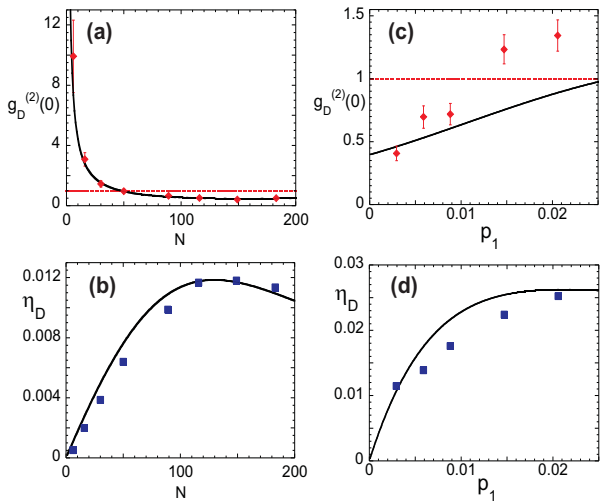,height=2.6in,width=3.1in}
\end{center}
\vspace{-0.5cm} \caption{ $g_D^{(2)}(0)$ as a function of maximum
number of trials $N$ (panel (a)) and $p_1$ (panel (c)); measured
efficiency to generate and detect a single photon $\eta_D$ as a
function of $N$ (panel (b)) and $p_1$ (panel (d)). For panels (a)
and (b) $p_1=0.003$ (about $6\cdot 10^5$ photons per {\it write}
pulse were used), whereas for for panels (c) and (d) $N=150$. The
full curves are based on Eq. (5) with the values of efficiencies and
coherence times given in the text, with however $\eta_D$ multiplied
by an empirical factor of $2/3$. We believe this reduced efficiency
is due to imperfect switching of the {\it read} light in the
feedback-based protocol (we note that there are no other adjustable
parameters in the simple theory presented). Evident deviations from
the theory in panels (c) and (d), beyond the statistical
uncertainties associated with photoelectric counting events, could
be explained either by inadequacies of the theory, or slow
systematic drifts in the residual magnetic field and the {\it read}
light leakage. } \label{TQ}
\end{figure}

In Fig.~3 we present the measured degree of 2nd order coherence for
zero time delay $g_D^{(2)}(0) \equiv P_{23}/(P_2P_3)$ \cite{loudon}
and the measured efficiency $\eta _D \equiv P_2 +P_3$ as a function
of $N$ (panels (a) and (b)), and as a function of $p_1$ (panels (c)
and (d)). The solid curves are based on Eq.(5). The dashed lines in
panels (a) and (c) show the expected value of $g_D^{(2)}(0)=1$ for a
weak coherent state (as we have confirmed in separate measurements).
The particular value of $\Delta t$ is chosen to optimize
$g_D^{(2)}(0)$ and $\eta _D$. The minimum value of
$g_D^{(2)}(0)=0.41 \pm 0.04$ indicates substantial suppression of
two-photon events and under the same conditions $\eta _D = 0.012$
\cite{Q}.  As shown in Fig. 3(a), when $N$ is small, the protocol
does not result in deterministic single photons. Instead, the
cleaning pulse-induced vacuum component of the idler field leads to
additional classical noise. Large $N$, and hence long coherence
times, are crucial to reduce this noise below the coherent state
level and to approach a single photon source. Note, that in the
limit of infinite atomic memory and $N\rightarrow \infty $,
$g_D^{(2)}(0) \rightarrow \mbox{min} \{\alpha \} \approx 0.012 \pm
0.007$ and $\eta _D \rightarrow \eta _i \approx 0.075$,
substantially exceeding the performance of any demonstrated
deterministic single photon source.

Moreover, $\eta _D$ can be further increased by employing atomic
sample with larger optical thickness and by optimizing the spatial
focusing patterns of the signal and idler fields \cite{effs}. In
principle, the spatial signal-idler correlations from an atomic
ensemble (and, therefore $\eta _i^0$) can also be improved by use of
an optical cavity. However, in the absence of special precautions
the use of a cavity will itself introduce additional losses
associated, e.g., with the mirror coatings or the cavity locking
optics \cite{kuhn,lange,black}. The measured efficiency $\eta _D $
would involve a trade-off between improved spatial correlations due
to the cavity and the concomitant losses that it introduces.

In conclusion, we have proposed and demonstrated a stationary source
of deterministic photons based on an ensemble of cold rubidium
atoms.

We thank M.S. Chapman for illuminating discussions. This work was
supported by the Office of Naval Research, National Science
Foundation, NASA, and Alfred P. Sloan and Cullen-Peck Foundations.


\begin{thebibliography}{99}
\bibitem{prob} This result can also be derived using
arguments based on elementary photon counting probabilities
\cite{gardiner}.
\bibitem{effs} In separate sets of measurements, we have
observed $\eta _s \approx 0.2$, for the intrinsic signal efficiency
$\eta _s^0 \approx 0.6$.
\bibitem{Q} The corresponding value of the measured Mandel
parameter $Q_D \equiv -\eta_D (1-g_D^{(2)}(0))$ is $ \approx -0.007
\pm 10\%$ and is largely determined by $\eta _D$ \cite{loudon}.
\end{thebibliography}
\end{document}